\documentclass{iucr}              % DO NOT DELETE THIS LINE
\usepackage {graphicx}
\usepackage {amsmath}
\usepackage {bm}

    \paperprodcode{a000000}      % Replace with production code if known
     \paperref{xx9999}            % Replace xx9999 with reference code if known
     \papertype{FA}               % Indicate type of article
     \paperlang{english}          % Can be english, french, german or russian
     \journalcode{A}              % Indicate the journal to which submitted
     \journalyr{2004}
     \journaliss{0}
     \journalvol{60}
     \journalfirstpage{000}
     \journallastpage{000}
     \journalreceived{0 XXXXXXX 0000}
     \journalaccepted{0 XXXXXXX 0000}
     \journalonline{0 XXXXXXX 0000}

\begin{document}                  % DO NOT DELETE THIS LINE

     %-------------------------------------------------------------------------
     % The introductory (header) part of the paper
     %-------------------------------------------------------------------------

     % The title of the paper. Use \shorttitle to indicate an abbreviated title
     % for use in running heads (you will need to uncomment it).

\shorttitle{SPEDEN}
\title{SPEDEN: Reconstructing single particles from their diffraction patterns}

\cauthor[a]{Stefan P. }{Hau-Riege}{E-mail: hauriege1@llnl.gov}

\author[a]{Hanna}{Szoke}
\author[a]{Henry N.}{Chapman}
\author[a]{Abraham}{Szoke}
\author[a]{Stefano}{Marchesini}
\author[a]{Alexander}{Noy}
\author[b]{Haifeng}{He}
\author[b]{Malcolm R.}{Howells}
\author[c]{Uwe}{Weierstall}
\author[c]{John C. H.}{Spence}

\shortauthor{S.~P.~Hau-riege et al.}

\aff[a]{Lawrence Livermore National Laboratory, USA.}
\aff[b]{Advanced Light Source, Lawrence Berkley National Laboratory, USA.}
\aff[c]{Department of Physics and Astronomy, Arizona State University, USA.}

\begin{synopsis}
SPEDEN is a computer program that reconstructs the electron density of 
single particles from their x-ray diffraction patterns, using an adaptation 
of the Holographic Method in crystallography. It is designed to deal 
successfully with sparse, irregular, incomplete and noisy data.
\end{synopsis}

\begin{abstract}
\textsc{Speden} is a computer program that reconstructs the electron 
density of 
single particles from their x-ray diffraction patterns, using a 
single-particle adaptation of the Holographic Method in crystallography. 
(Szoke, A., Szoke, H., and Somoza, J.R., 1997. Acta Cryst. A53, 291-313.) 
The method, like its parent, is unique that it does not rely on ``back'' 
transformation from the diffraction pattern into real space and on 
interpolation within measured data. It is designed to deal successfully with 
sparse, irregular, incomplete and noisy data. It is also designed to use 
prior information for ensuring sensible results and for reliable 
convergence. This article describes the theoretical basis for the 
reconstruction algorithm, its implementation and quantitative results of 
tests on synthetic and experimentally obtained data. The program could be 
used for determining the structure of radiation tolerant samples and, 
eventually, of large biological molecular structures without the need for 
crystallization.
\end{abstract}


\section{Introduction}

This paper describes our computer program \textsc{Speden} that reconstructs 
the density from the diffraction patterns of individual particles. 
\textsc{Speden}
is of interest for three reasons. Diffractive imaging promises to 
improve the resolution, sensitivity, and practical wavelength range in X-ray 
microscopy, for three-dimensional objects that are tolerant to X-rays. A few 
examples are defects in semiconductor structures, phase separation in 
alloys, nano-scale machines and laser fusion targets. A long-term vision is 
the possibility of high-resolution reconstruction of diffraction patterns of 
single bio-molecules. Of broad theoretical interest is \textsc{Speden}'s 
unique approach to the reconstruction of scatterers - a difficult 
mathematical problem. In the rest of this section we expand on these three 
topics.

Reconstruction of the electron density from non-uniformly sampled, 
three-dimensional diffraction patterns is of wide interest and applicability 
with present-day sources. In radiation-tolerant samples, x-ray diffraction 
and diffraction tomography are capable of higher resolution than (straight- 
or cone-beam) tomography alone. In tomography, resolution is limited by the 
quality of the incident beam and by the spatial resolution of the detector; 
in diffraction the resolution can be as fine as the wavelength of the 
incident radiation. Experimentally, diffraction imaging has already produced 
X-ray images at higher resolution than possible with available X-ray optics 
(Miao et al., 1999, and He et al., 2002). The price to be paid for these 
benefits is the intrinsic difficulty of the reconstruction. Nevertheless, 
several successful reconstructions from experimental X-ray data, using the 
iterative hybrid input-output version of the Gerchberg-Saxton-Fienup (GSF) 
algorithm, have been reported recently (Miao et al., 2002, and Marchesini et 
al., 2003). The first successful application of this algorithm to electron 
diffraction data was reported in 2001 (Weierstall et al., 2002), and it has 
been used more recently to produce the first atomic-resolution image of a 
single carbon nanotube (Zuo et al., 2003). In biology, the use of the GSF 
has recently been shown to dramatically reduce the number of images needed 
for tomographic cryoelectronmicroscopy of protein monolayer crystals, so 
that phasing can be based mainly on the three-dimensional diffraction data 
(Spence et al., 2003). 

The development of \textsc{Speden} was also prompted by the promise of new 
ways to image bio-molecules. Free-electron lasers can, in principle, provide 
X-ray pulses of tens to hundreds of femtoseconds in length and brightness up 
to ten orders of magnitude greater than synchrotron radiation. It was 
predicted that, under such circumstances, it should be possible to dispense 
with crystals and reconstruct the electron density of single biological 
particles from their diffraction pattern (Neutze et al., 2000). In proposed 
experiments, a large number of single particles will be injected into the 
X-ray beam in random orientation and their diffraction patterns will be 
recorded, each in a single shot of the free-electron laser. Such diffraction 
patterns will be very noisy and their resolution will be limited by the 
signal to noise (S/N) ratio. The measured diffraction patterns that 
correspond to different orientations of the particle will be classified into 
a number of mutually exclusive classes. The images within each class will 
then be averaged and the class averages assembled into a three-dimensional 
diffraction pattern by finding their mutual orientation relationships. 
Finally, the three-dimensional diffraction pattern will be reconstructed to 
yield the electron density of the molecule.

We have worked on the analysis of all three steps of such an experiment. The 
essence of the first analysis is that the maximum X-ray intensity at a given 
pulse length is limited by the requirement that the molecule stay intact 
during the pulse, even though it eventually disintegrates (Hau-Riege et al., 
2003). The second analysis discusses the division of noisy diffraction 
patterns into a number of distinct classes. If the images are divided into 
too few classes, the available resolution is not realized. If the patterns 
are divided into too many classes, the class averages will be poor and the 
pattern quality suffers. The individual class averages, each corresponding 
to a well-defined orientation of the particle, will be assembled into a 
three-dimensional diffraction pattern. The result will be a 
three-dimensional diffraction pattern that is measured at a limited number 
of orientations. It will be, therefore, sparse, irregular and will have a 
limited signal to noise ratio (Huldt et al., 2003).

The program \textsc{Speden}, described in this paper, provides a way to 
optimally determine the electron density from such a three-dimensional 
ensemble of continuous diffraction patterns. We first give an analysis of 
their properties and discuss the methods and the expected difficulties of 
reconstructing a ``sensible'' electron density from them. We then describe 
how \textsc{Speden} adapts the Holographic Method (Szoke et al., 1997a) in 
crystallography to deal with continuous diffraction patterns as opposed to 
discrete Bragg spots; this will be discussed in the next section. We then 
report quantitative results of preliminary tests for verifying the 
correctness of our method. These tests use computed and measured diffraction 
patterns from samples of inorganic particles.

\section{SPEDEN: The Program}

\subsection{Theoretical Considerations}

\subsubsection{Mathematical background}

The reconstruction of the density of scatterers from its diffraction pattern 
is an ``inverse problem''. Other, well-studied inverse problems are those of 
computed tomography, image deblurring, phase recovery in astronomy, and 
crystallography. In tomography, for example, an inversion algorithm (e.g. 
filtered back projection) is used to recover the density of scatterers from 
the measured tomograms. It is widely recognized that the reconstructed 
density is very sensitive to inaccuracies in the measurement. Small errors 
in the diffraction pattern cause large errors in the reconstruction. This 
property is called ill-posedness or ill conditioning. 

The reconstruction of the electron density from x-ray diffraction patterns 
is indeed ill conditioned. It also has two \textit{additional} difficulties. First, in contrast 
to tomography, there are no direct inversion algorithms -- not even 
approximate ones. Second, the reconstructed electron density at any sample 
point is influenced strongly by the electron densities of all sample points, 
as opposed to a limited number of them. Therefore, errors in density are 
non-local and ``propagate'' far. 

Fortunately, very good fundamental discussions of these subjects are 
provided in the books of Daubechies (Daubechies, 1992), Bertero and Boccacci 
(Bertero et al., 1998, Bertero, 1989) and Natterer (Natterer, 1996, Natterer 
and W\"{u}bbeling, 2001). In somewhat simplified terms, the reconstruction 
of the electron density is similar to finding the inverse of an 
ill-conditioned non-square matrix, a subject thoroughly discussed in Golub 
and van Loan (Golub et al., 1996). We consider these mathematical properties 
to be essential for understanding the successes and limitations of 
reconstruction algorithms; we will try to be fully cognizant of them in the 
discussion that follows.

\subsection{The phase problem of crystallography, oversampling}

The crystallographic phase problem is a good starting point for further 
discussion. It was first realized by Sayre (Sayre, 1952) that the number of 
observable complex structure factors, limited by the Bragg condition, is 
equivalent to a critical sampling of the electron density in the unit cell 
of the crystal. The sampling theorem of Whittaker and Shannon teaches us 
that, if the amplitudes and phases of all the diffraction peaks were 
accurately measured, the electron density could, in principle, be 
reconstructed everywhere (Bricogne, 1992). Unfortunately, only the 
amplitudes of the Bragg reflections are measured, not their phases. 
Therefore there is not enough information in the diffraction pattern for a 
unique reconstruction of the electron density. Sayre proposed (Sayre, 1980) 
that if we could measure the diffraction amplitudes ``in between'' the Bragg 
peaks, we should have enough information to reconstruct the electron 
density, or to ``phase'' the diffraction pattern. This is exactly the 
situation in diffraction from a single particle. 

Nevertheless, it is still difficult to reconstruct the electron density, 
even from a well ``oversampled'' diffraction pattern. One corollary of 
critical sampling is that the amplitudes and phases of the Bragg reflections 
of a crystal are independent of one other, but any structure factor in 
between them depends on the surrounding ones to some extent. Therefore, too 
much oversampling does not help to obtain independent data, although it does 
improve the S/N ratio by reducing the noise. Ill posedness is still with us, 
although with oversampling, the error propagates less. An additional 
difficulty with diffraction patterns from a set of discrete orientations of 
a particle is that at low resolution the diffraction pattern is well 
oversampled while at high resolution the sampling is sparse. A fundamental 
property of diffraction is that the position and the handedness of the 
electron density are undetermined, resulting sometimes in stagnation of the 
algorithm and drift in the position of the results (Stark, 1987). 

There are two well-known, necessary remedies for the lack of information and 
for the ill posedness of the reconstruction problem. The more important one 
is the need for more information. For example, one way to include \textit{a priori} knowledge 
is to accept reconstructed electron densities only if they are 
``reasonable''. The second remedy is to use ``stabilized'' or 
pseudo-inversion algorithms. In the next section we introduce our version of 
a real space reconstruction algorithm; we will argue that our algorithm 
deals with all these problems optimally, at least in some sense. We return 
to the comparison of our algorithm with other methods for phase-recovery in 
Section 2.3.

subsection{Speden, a real-space algorithm}

In this section we outline the workings of our reconstruction program, 
\textsc{Speden}. \textsc{Speden} uses a real-space method for 
reconstruction; its acronym stands for Single Particle Electron DENsity. For 
computational efficiency the particle to be recovered is put into a 
\textit{fictitious} unit cell that is several times larger than the 
particle itself. All 
reconstruction algorithms use this artifice in order to be able to calculate 
structure factors by fast Fourier transform techniques. The resulting 
similarity with crystallography enables the use of many crystallographic 
concepts. In fact, the recognition of this similarity enabled us to write a 
program, \textsc{Speden}, based on our crystallographic program, 
Eden, with relatively small modifications.

The most significant difference between the two programs is that in 
crystallography the Bragg condition restricts the reciprocal lattice vectors 
to integer values, while the continuous diffraction pattern can be - and 
usually is - measured at arbitrary, non-integer values of the reciprocal 
lattice vectors.

In \textsc{Speden}, in common with \textsc{Eden}, the (unknown) electron 
density is represented by a set of Gaussian basis functions, with unknown 
amplitudes, that fill the fictitious unit cell uniformly. This way the 
recovery is reduced to the solution of a large set of quadratic equations. 
The program ``solves'' these equations by finding the number of electrons in 
each basis function so as to agree optimally with the measured diffraction 
intensities as well as with other ``prior knowledge''. Prior knowledge 
includes the emptiness of the unit cell outside the molecule, the positivity 
of the electron density, possibly some low-resolution image of the object, 
etc. Each one of those conditions is described by a cost function that 
measures the deviation of the calculated data from the observed data. One of 
the cost functions describes the deviation of the calculated diffraction 
pattern from the measured one; others depend on the deviation of the 
recovered density from prior knowledge. Measured data are weighted by their 
certainty (inverse uncertainty), other prior knowledge is weighted by its 
``reliability''. The mathematical method used is (constrained) conjugate 
gradient optimization of the sum of cost functions. At each step of the 
optimization, there is a set of amplitudes available that describe the 
current electron density in the full unit cell. A full set of structure 
factors is calculated by Fourier transforming the current electron density. 
When the unit cell is larger than the particle, the structure factors can be 
stably interpolated to compare them with measured structure factor 
amplitudes. 

We refer to the cited literature that shows that the procedure we outlined 
is equivalent to a stabilized (quasi) solution of the inverse problem 
(Daubechies, 1992; Bertero et al., 1998; and Natterer et al., 2001). As 
such, it is optimally suited for sparse, irregular, incomplete and noisy 
data. 

In the following subsections we describe very briefly the common features of 
Eden and Speden as well as their differences. A more 
complete description of Eden can be found in previous papers (Szoke 
et al., 1997a, and Szoke, 1998).

\subsubsection{Representation of the electron density}

The electron density is represented as a sum of basis functions, adapted to 
the resolution of the data. Specifically, we take little Gaussian ``blobs'' 
of width comparable to the resolution, and put their centers on a regular 
grid that fills the ``unit cell'' and whose grid spacing is comparable to 
the resolution. The amplitudes of the Gaussians are proportional to the 
local electron density. In fact, the number of electrons contained in each 
Gaussian constitutes the set of our basic unknowns. The above is identical 
to the representation of the electron density in \textsc{Eden.}

Some mathematical details follow. The actual formula for the representation 
of the electron density as a sum of Gaussians is

\begin{equation}
\label{eq1}
\rho _\text{unknown} ({\rm {\bf r}}) \simeq \frac{1}{(\pi \eta \Delta r^2)^{3 / 
2}}\sum\limits_{p = 1}^{P} {n(p)} \exp \left[ {\frac{ - \vert {\rm {\bf r}} 
- {\rm {\bf r}}(p)\vert ^2}{\eta \Delta r^2}} \right].
\end{equation}

The centers of the Gaussians are positioned at grid points, $\bm{r}(p)$, 
where $p$ is a counting index. In our fictitious unit cell, the grid is 
orthogonal, the grid spacing is $\Delta r$ and the centers of the Gaussians 
are usually on two intercalating (body centered) grids for best 
representation of the electron density. The number $\eta$ of 
the order unity, determines the width of the Gaussians relative to 
their spacing, $\Delta r$. Finally and most importantly, $n(p)$ 
is the \textit{unknown} number of 
electrons in the vicinity of the grid point $\bm r(p)$. The 
values of $n(p)$ are real, and in future may also be complex-valued 
to allow for photo-absorption in addition to scattering. 
(The latter can be significant 
when diffraction measurements are made at longer X-ray wavelengths.)

Given $n(p)$ the structure factors can be calculated by
\begin{equation}
\label{eq2}
F_{calc} (\bm{h}) = 
e^{ %\left [ 
	- \eta \left (
		\pi \Delta r \left | 
			{\cal F}^T \bm{h}
			\right | 
		\right )^2
	%\right ]
}
\sum\limits_{p = 1}^{P} n(p)\exp \left [
	2\pi i \bm{h} \cdot {\cal F}\bm{r}(p)
	\right ]\,,
\end{equation}
using a fast Fourier transform. The vector $\bm {h}$, a triplet of 
integers, denotes the reciprocal lattice vector, the operator F transforms 
from real space (Cartesian) coordinates to fractional coordinates, and 
$\cal{F}^{T}$ denotes the dual transformation.

The constants appearing in Eqs. (\ref{eq1}, \ref{eq2}) were discussed in some detail 
previously (Szoke et al., 1997b). For completeness, we define them here. The 
crystallographic $B$ factor is given by $B  = \left( {2\pi \Delta r} 
\right)^2\eta $. The ``crystallographic resolution'', $d$, determines the grid 
spacing, $\Delta r$ by the relation, $\Delta r  \approx  f_\text{space} d$, 
where $f_\text{space}$ is a constant of the order unity. For a body-centered 
lattice we set $f_\text{space}$ = 0.7 and $\eta $ = 0.6. 
For a simple lattice, we 
use $f_\text{space}$ = 0.6 and $\eta $ = 0.8.

Note that the Gaussian basis functions are not used in a one-to-one 
correspondence with single atoms, but are simply used to describe the 3D 
electron density at the resolution that is appropriate to the data 
resolution. In the special case that the resolution was about the size of an 
atom and an atom happened to be sitting exactly on a grid point, that atom 
would be represented by a single basis function. If the atom is not on a 
grid point, or if the atom happens to be fat, because of thermal motion, 
that same atom would be represented by many basis functions. Similarly, at 
lower resolution, a single basis function represents assemblies of atoms.

\subsubsection{Reciprocal space knowledge}

The measured diffraction pattern of the molecule is proportional to the 
absolute square of the structure factors. In \textsc{Speden} we do account 
for the curvature of the Ewald sphere. There are two subtle points: the 
diffraction pattern is measured only in a finite number of directions, 
$\bm H(i)$; and as a rule, those directions are not along the reciprocal 
lattice vectors of the (fictitious) unit cell for a single particle. In 
other words, the measurement directions, $\bm H(i)$, are usually not 
integers and they are not uniformly distributed in reciprocal space. This is 
the main difference between crystallography and single particle diffraction 
and, therefore, between \textsc{Eden} and \textsc{Speden}. The essence of 
\textit{any} reconstruction algorithm is to try to find an electron density distribution 
such that the calculated diffraction pattern matches the observed one. In 
our representation, we try to find a set of $n(p)$, such that 

\begin{equation}
\label{eq3}
\left | F_\text{obs} (\bm H(i))\right | ^2 = 
\left | F_\text{calc}(\bm H (i))\right | ^2 \,,
\end{equation}
for each measurement direction, $\bm{H}(i)$. Let us assume for a moment 
that $\bm{H}(i)$ are integers. When the representation of the unknown 
density is substituted from (\ref{eq2}), for each measured value of 
$\bm{H}(i)$, equation (\ref{eq3}) becomes a quadratic equation in the 
unknowns, $n(p)$. The number of equations is the number of 
measured diffraction 
intensities. It is usually not equal to the number of independent unknowns 
that are the number of grid points in the unit cell. The equations usually 
contain inconsistent information, due to experimental errors. The equations 
are also ill conditioned and therefore their solutions are extremely 
sensitive to noise in the data. Under these conditions the equations may 
have many solutions or, more usually, no solution at all. Our way of 
circumventing these problems is to obtain a ``quasi solution'' 
of (\ref{eq3}) by minimizing the discrepancy, or cost function 
(see e.g. Stark, 1987, and Bertero et al., 1998)

\begin{equation}
\label{eq4}
\chi _{Speden} = \sum\limits_{i} {w'(\bm{H}(i))^2} 
\left[ 
\left| 	F_{obs} (\bm{H}(i)) \right| - 
\left| F_{calc} (\bm{H}(i)) \right| 
\right]^2
\end{equation}

The weights, $w'(\bm H(i))^{2}$, are usually set to be proportional to 
the inverse square of the uncertainty of the measured structure factors, 
$1/\sigma ^{2}(\bm{H}(i))$. As discussed by Szoke (Szoke, 1999), this 
is equivalent to a maximum likelihood solution of the equations. 

Let us now discuss the first, previously ignored difficulty in the 
reconstruction. When we try to reconstruct the electron density from real 
experimental data, we have to compare the set of measured 
$| F_\text{obs}(\bm{H}(i)) |$, 
where $\bm{H}(i)$ are not necessarily integers, with the calculated 
structure factor amplitudes, $|F_\text{calc} (\bm{h})|$, 
that are on a regular grid, i.e. have 
integer $\bm{h}$. In principle, given an electron density of the molecule, 
one could calculate the structure factors in the experimental directions. 
Nevertheless, for computational efficiency, we put the (unknown) molecule or 
particle into a fictitious unit cell that is larger than the molecule. We 
will also demand that the Gaussians outside the molecular envelope be empty. 
(In practice, sizes of molecules are known from their composition; particle 
sizes and shapes may be known from lower-resolution imaging.) As long as the 
distances of the Gaussian basis functions are kept to be the experimental 
resolution, the number of ``independent'' unknowns neither increases nor 
decreases, in principle, by this computational device. The structure factors 
are calculated on an integer grid in the large unit cell, so they are 
essentially oversampled in each dimension by the same factor of the size of 
the large cell to the size of the molecule. The oversampling allows stable 
interpolation of the calculated structure factor amplitudes from integer 
$\bm{h}$ to the fractional $\bm H(i)$ everywhere, independent of the 
density of the actual measurements. Note that interpolation from fractional 
$\bm H(i)$ to integer $\bm{h}$ is not always a stable procedure!

In the present implementation of \textsc{Speden}, we get sufficient accuracy 
with the simplest, tri-linear interpolation in the amplitudes of $\vert 
F_\text{calc}(\bm{h}) \vert $ if we choose the fictitious unit cell to be 
three times larger than the molecule in each dimension. Now, some 
mathematical details: the reciprocal space vector \textbf{H}($i)$ is within a 
cube, bounded by eight corners $\bm{h}(i,j),\,, \{j=1,8\}$ with integer 
values. Let us denote the fractional parts of the components of 
$\bm{H}(i)$ as $(H, K, L)$. We define weights for the eight corners, $w(i,j)$, by taking the 
products of the fractional parts of $H$, or $(1-H)$ with those of $K$ or $
(1-K)$ and $L$ or 
$(1-L)$. The cost function to be minimized now becomes

\begin{eqnarray}
\label{eq5}
\nonumber
\chi _\text{Speden} &=& \sum\limits_{i} w'({\rm {\bf H}}(i))^2 \cdot\\
&& \left[ {\vert 
F_{obs} ({\rm {\bf H}}(i))\vert - \sum\limits_{j = 1}^{8} {w(i,j)} \vert 
F_{calc} ({\rm {\bf h}}(i,j))\vert } \right]^2\,.
\end{eqnarray}

A similar approach of applying crystallographic algorithms to continuous 
diffraction data has been done with direct methods [Spence et al., 2003]. In 
this case, however, the Ewald sphere was approximated by a plane.

\subsubsection{Real space knowledge (targets)}

Let us assume that we have some, possibly uncertain, knowledge of the 
electron density in parts of the unit cell from an independent source, i.e. 
one that does not come from the X-ray measurement itself. This is the kind 
of knowledge present when the unknown molecule is placed into a larger unit 
cell and we demand that the unit cell be empty outside the molecule. This 
kind of knowledge was also referred to as a ``sensible'' electron density in 
the introduction. We represent this knowledge by a target electron density 
$n_\text{target}(p)$ and by a real-space weight function $w'(p)^{2}$. 
It will be 
desirable that the actual electron density of the molecule, $\rho 
(\bm{r}) $, as represented by $n(p)$, be close to the target 
electron density; the weight function $w'(p)^{2}$ expresses the strength of 
our belief in the suggested value of the electron density. Note that target 
densities can be assigned in any region of the unit cell independently of 
those in any other region. The simplest way to express the above statement 
mathematically is to minimize the value of the cost function

\begin{equation}
\label{eq6}
\chi _\text{space} = 
\frac{A}{2}\lambda_\text{space} 
\sum\limits_{p=1}^{P} 
w'(p)^2 \left [ n(p) - n_\text{target}(p)  \right]^2 \,,
\end{equation}

\noindent
where $\lambda _{space}$ is a scale factor and $A$ is a normalizing constant, 
described in Somoza et al., 1995. (In the \textit{absence} of 
information at and around the 
molecule, weights are generally unity where it is known that there is no 
molecule and zero elsewhere.) The same procedure is used in \textsc{Eden.}

\subsubsection{Low resolution target (phase extension)}

The knowledge of the electron density at low resolution can be expressed by 
a low-resolution spatial target. Crystallographers call this phase 
extension. The essence is that, during the process of the search for an 
optimal electron density, we try to keep its low resolution component as 
close to the known density as possible. A convenient way to accomplish this 
is the following. Given $n(p)$, we smear out its Gaussian representation and 
compare it to the equally smeared out target. Actually, it is easier to 
carry out the computation in reciprocal space. We define
\begin{equation}
\label{eq7}
\chi_\text{phasext} = \lambda_\text{phasext} 
\sum\limits_{i} w' (\bm{h})^2 
\left |
	 F_\text{smear} (\bm{h}) - F_\text{tar} (\bm{h})
\right |^2,
\end{equation}
where the current ``smeared'' structure factors are calculated using the low 
resolution, $\Delta R$,
\begin{equation}
\label{eq8}
F_{\text{smear}} ({\rm {\bf h}}) = \exp [ - \eta (\pi \Delta R\vert 
{\cal F}^T{\rm {\bf h}}\vert )^2]\sum\limits_{p = 1}^{P} {n(p)\exp [2\pi 
i{\rm {\bf h}} \cdot {\cal F}{\rm {\bf r}}(p)]} .
\end{equation}

The original low-resolution target is prepared analogously from the 
(presumably) known electron density, $n_\text{tar}(p)$. The same procedure is used 
in \textsc{Eden.} 

\subsubsection{Minimization of the cost function}

In the presence of a target density, the actual cost function used in the 
computer program is the sum of $\chi _\text{Speden}$ (\ref{eq5}), $\chi _\text{space}$ (\ref{eq6}) and $\chi _\text{phasext}$ (\ref{eq7})

\begin{equation}
\label{eq9}
\chi_\text{total} = \chi_\text{Speden}+ \chi_\text{space} + 
\chi_\text{phasext}\, .
\end{equation}

The fast algorithm described in Somoza et al., 1995, and Szoke et al., 
1997b, is always applicable to the calculation of the full cost function, 
(\ref{eq9}). There is a clear possibility of defining more target functions. They 
are all added together to form $\chi _\text{total}$ that is minimized to find 
the optimum electron density.

The minimization of the cost function (\ref{eq9}) is carried out in \textsc{Speden} 
(as in \textsc{Eden}) by D. Goodman's conjugate gradient algorithm (Goodman, 
1991). It has proven to be very robust and efficient in years of use in 
\textsc{Eden}. The essential properties of the algorithm that make it so 
advantageous for our application is that the positivity of the electron 
density, $n(p)\ge  0$, is always enforced and that the gradient vector in 
real space can be calculated by fast Fourier transform. The gradient 
calculation needed only a very simple modification for the interpolation in 
reciprocal space, Eq. (\ref{eq5}). The line search algorithm does not use the 
Hessian, so matrices are never calculated.

As with any local minimization, global convergence is not achieved. We 
discussed this problem in our previous papers and came to the conclusion 
that, usually, the minimum surface of the cost function (\ref{eq7}) is so 
complicated that finding a global minimum would take more computer time than 
the existence of the universe. 

\subsection{Comparison to iterative algorithms}

Reconstruction of the scatterer from a continuous diffraction pattern has a 
tangled history replete with repeated discoveries. Some of the present 
authors are also guilty of ignorance of prior work. We referred to the 
pioneering insights in Section 2.1.2. 

The ``recent'' period of algorithms started with the work of Miao, Sayre and 
Chapman (Miao et al. 1998) who pointed out that the fraction of the unit 
cell where the density is known is an important parameter for convergence. 
In somewhat later work, with oversampled structure factors calculated on a 
regular grid, the crystallographic program \textsc{Eden} successfully 
demonstrated the recovery of the electron density using a simulated data set 
from the Photoactive Yellow Protein (Szoke, 1999). The protein was put into 
a fictitious unit cell, twice the size of the original one, and a target 
with zero density was used outside the original unit cell of the protein. 
Similarly, Miao and Sayre (Miao et al., 2000) have studied empirically how 
much oversampling is required in two- and three-dimensional reconstructions 
of a simulated data set; using a version of the Gerchberg -- Saxton -- 
Fienup (GSF) algorithm. Among recent articles we mention Robinson et al 
(2001), Williams et al (2003), Marchesini et al. (2003) and references 
therein, in addition to those mentioned in the Introduction.

All reconstruction algorithms of oversampled diffraction patterns use \textit{a priori} 
information on the shape and size of the particle. In our previous studies 
in crystals we found that such information is very valuable. For example, 
\textsc{Eden} converges surprisingly well for proteins at low resolution 
where the only information used is that the molecule is a single ``blob''. 
\textsc{Eden} also converges for synthetic problems with a good knowledge of 
the solvent volume, which is greater than 50{\%} (Beran et al., 1995) or 
60{\%} (\textsc{Eden}). A similar conclusion was reached in Miao et al. 
(1998). In comparison, when a molecule is embedded in a 3-fold larger 
fictitious unit cell, the empty ``solvent'' occupies $\sim $96{\%} of the 
cell volume. 

As discussed previously, the reconstruction of scatterers from their 
diffraction pattern is a difficult mathematical problem. In many cases the 
indeterminacy of the absolute position of the object and of its handedness 
causes difficulties in convergence. That is definitely the case with 
\textsc{Speden} so, in that sense, \textsc{Speden} is not a good algorithm. 
Empirically, the GSF algorithm has a larger radius of convergence and deals 
better with stagnation (Marchesini et al. 2003)

Another family of difficulties arises when there is \textit{a priori} information available, 
but there is only incomplete and noisy data. Under such conditions the main 
questions are how to find a solution that optimally takes into account the 
available information and that is the best ``sensible'' one that reproduces 
the noisy and incomplete data to its limited accuracy. It is this second set 
of conditions for which \textsc{Speden} was written. Although, in this 
paper, we show only its performance for artificial and ``easy'' but 
incomplete data, \textsc{Speden}'s older sister, \textsc{Eden} has been 
shown to have those properties on a large range of crystallographic data 
sets, ranging from CuO$_{2}$ to the ribosome. We expect that such properties 
of \textsc{Eden} will be inherited by \textsc{Speden}, considering that 
their fundamental mathematical properties are sufficiently similar.

The best known, and successful class of algorithms is the group of iterative 
transform algorithms that we refer to as Gerchberg-Saxton (GS) (Gerchberg et 
al., 1972), and its development, in which support constraints and feedback 
are added, the GS-Fienup (GSF) or hybrid input-output algorithm (Fienup et 
al., 1982; Aldroubi et al., 2001; Bauschke et al., 2002; Bauschke at al., 
2003). The essence of the GSF algorithms is that they iterate the N-pixel 
data between real and reciprocal spaces via FFTs and enforce the known 
constraints in each of these spaces. 

Depending on the degree of noise in the data, these algorithms usually 
converge in about a hundred to a thousand iterations. The weak convergence 
(non divergence) of the GS algorithm has been proven in the absence of noise 
(Fienup et al., 1982). There is no mathematical proof that these algorithms 
will converge in general, but it is reasonable that by sequentially 
projecting onto the set that satisfies the real space constraints and the 
set that satisfies the reciprocal space constraints, the intersection 
(corresponding to a valid solution) should be approached. This is definitely 
true for projections onto convex sets, but unfortunately these sets are not 
convex (Stark, 1987). In practice, despite the fact that the modulus 
constraint is non-convex, the algorithms often converge even in the presence 
of noise in about several hundred iterations.

The main difference between the GSF algorithms and S\textsc{peden} is that 
\textsc{Speden} does not iteratively project onto the sets of solutions that 
satisfy the real-space or reciprocal space constraints separately, but 
rather it minimizes a cost function that includes all the constraints of 
both spaces. It does this by varying quantities in real space only (the 
$n(p)$'s) and the cost evaluation only requires a forward transform from real 
space to reciprocal space. As the cost function never increases, 
\textsc{Speden} reaches only a local minimum.

In spite of its ``small'' radius of convergence, there are some expected 
advantages to \textsc{Speden}'s algorithm.. In \textsc{Speden} we compare 
the $F_{calc}$ to the $F_{obs}$ by interpolating from the samples of 
$F_{calc}$(\textbf{h}), calculated on a regular grid, to the measured sample 
vectors \textbf{H}. Since the gridded $\vert F_\text{calc}(\bm{h})\vert $ 
are a complete set (due to the fact they are sampled above the Nyquist 
frequency) the interpolation is stable and performed with little error. 
Since an inverse transform is required in the GSF algorithms, the measured 
diffraction data $\vert F_{obs}(\bm{H}(i))\vert $ (recorded on Ewald 
spheres in reciprocal space) must be interpolated to the gridded data points 
\textbf{h}. The observed data might not be a complete set, especially at 
high resolution where the density of samples is sparser: this may lead to 
error. An additional possible difficulty with the GSF algorithms is that the 
effective number of unknowns may increase with the size of the fictitious 
unit cell, while in \textsc{Speden} the effective number of unknowns is 
constant. Finally, in \textsc{Speden}, weightings can be properly applied to 
all data and knowledge. The measured data is inversely weighted by its 
uncertainty; it is a procedure equivalent to maximum likelihood methods and 
it should be optimal, at least in theory. As a side effect, when the 
constraints are inconsistent, \textsc{Speden} still converges to a 
well-defined and correct solution. (Note that for noisy data the constraints 
are almost always inconsistent.) Weightings are also applied to reflect our 
confidence in real-space constraints, and these weightings are consistently 
used in real space.

\textbf{Tests}

\textsc{Speden} has certain built-in limitations. In particular, of course, 
reconstruction is only as good as the diffraction measurements and derived 
structure factor amplitudes are reliable. There are also other less obvious 
limitations. For example, there are inherent inaccuracies due to the 
Gaussian representation of the electron density in real space. (Szoke et 
al., 1997a) Also, the trilinear interpolation for representing integer (hkl) 
structure factor amplitudes on a non-integer grid is only approximate. 
Finally, as a fundamental limitation of \textit{any} reconstruction method, both the 
absolute position and the handedness of the molecule are undefined.

We performed preliminary tests to verify the capabilities and limitations of 
our reconstruction method using computed and experimentally obtained 
diffraction patterns. In this section, we first describe the reconstruction 
of simple ``molecules'' from synthetic diffraction patterns with 
\textsc{Speden}. Specifically, we discuss how the convergence of 
\textsc{Speden} is affected by the errors due to interpolation in reciprocal 
space, by the quantity of observed structure factors, $\vert 
F_\text{obs}\vert $, by the extent of the ``known'' starting model, and by 
the uniformity of sampling in reciprocal space. We then describe the 
reconstruction of simple two-dimensional objects from synthetic and measured 
diffraction patterns with \textsc{Speden}.

\subsection{Generation of Synthetic Test Data}

We created synthetic ``molecules'' in the format of the Protein Data Bank 
(pdb) files (Berman et al., 2000). Each molecule was composed of 15 
point-like carbon atoms, placed at random positions within a cube measuring 
16.8 {\AA} in each dimension and ``measured'' to 4 {\AA} resolution; these 
values correspond to crystallographic $B$ factors of 185.7 {\AA}$^{2}$. The 
molecule was then shifted so that its center-of-mass was at the center of 
the cube. All our simulations were repeated using molecules with several 
different random arrangements. Initially, the ``unit cell'' coincided with 
the dimensions of the cube in which the atoms were placed; later, larger 
cells were used and the atoms were positioned in their center. We also 
generated ``starting models'' by removing atoms from the full ''molecule''.

Both full and partial molecules served to generate structure factors 
($F_\text{calc}$), 
using the atomic positions and the B factors. Starting models 
were generated from the $F_{calc}$, using \textsc{Speden}'s preprocessor, 
\textsc{Back}, which finds the optimal real-space representation for an 
input $F_\text{calc}$. Initially, sets of ``measurements'' ($|F_\text{obs}|$) were generated by deleting the phases of the calculated 
structure factors of the full or partial molecule. The $F_\text{obs}$ files so 
generated had all integer $\bm H(i)$. The uncertainty of the measured 
structure factors, $\sigma(\bm{h})$, were chosen to be

\begin{equation}
\label{eq10}
\sigma \left( {\rm {\bf h}} \right) = \alpha \sqrt {\left\langle {F_{obs} 
({\rm {\bf h}}')} \right\rangle _{h'} / 10 + F_{obs} ({\rm {\bf h}})} ,
\end{equation}

\noindent
with $\alpha = 0.1$ for $\bm{h} \ne  (0,0,0)$, and $\alpha  = 0.01$ 
for $\bm{h} = (0,0,0)$. We used two alternate methods to generate 
$F_\text{obs}$ files for non-integer $\bm{H}(i)$: 
In the first method, we used a 
unit cell whose dimensions were incommensurate with one another and with the 
edge of the cube, but whose volume equaled that of the cube. The resulting 
$F_\text{obs}$ file, generated again from $F_\text{calc}$ files 
by deleting the 
phases, then had its indices scaled back appropriately, yielding fractional 
$\bm{H}(i)$. A different second method was used to sample the reciprocal 
$\bm{H}(i)$ space non-uniformly: $|F_\text{obs}| $ at fractional 
$\bm{H}(i)$ were calculated using tri-linear interpolation from 
$|F_\text{obs}|$ data on a regular grid that, in turn, was at four times 
the regular resolution in reciprocal space.

For each of these $F_\text{obs}$ files, some constraining 
information is required 
in order to find the atom positions. We used two types of constraints: one 
of them identified the (approximate) empty region; the other one used 
$F_\text{calc}$ at a considerably lower resolution. We call them the empty target 
and the low-resolution target, respectively. Both are based on the 
assumption that at a considerably lower resolution, the general position of 
atoms as one or more ``blobs'' in empty space is known. The low-resolution 
$FS_\text{calc}$ was prepared by smearing the full $F_\text{calc }$ 
file to 10 {\AA}. 
The empty target identified the empty points in terms of a mask. Then, 
throughout \textsc{Speden}'s iteration process, using the solver 
\textsc{Solve}, the program attempted to match the current electron/voxel 
values in masked-in regions to empty ($n(p)=0$) values. 
The phase-extension target 
used the same $FS_\text{calc}$ file; 
during the iteration process, at each step, 
the current real-space solution was smeared to that low resolution in 
reciprocal space and restrained to agree with that target.

\subsection{Assessment of Quality of Reconstruction}

Besides inspecting the reconstructed electron density visually, we used four 
quantitative measures to compare the reconstructed image with the electron 
density from the full molecule at 4 {\AA}:
\begin{enumerate}
\renewcommand{\theenumi}{\alph{enumi}}
\item
 Real-space RMS error: We calculated the real-space electron density from 
the electron/voxel files, a process we call regridding. We then calculated 
the root-mean-square (RMS) error of the electron densities, $\rho _{{\rm 
x}}$ and $\rho _{2}$, defined as
\begin{equation}
\label{eq11}
RMS = \frac{\sqrt {\sum\limits_{r} {\left( {\rho _{1} (r) - \rho _{2} (r)} 
\right)^2} } }{\sqrt {\sum\limits_{r'} {\rho _{1}^{2} (r') + } 
\sum\limits_{r"} {\rho _{2}^{2} (r")} } },
\end{equation}
We permitted one file to be shifted with respect to the other file, in order 
to minimize the distance.

\item Phase difference: We calculated the average (amplitude-weighted) phase 
difference between the $F_\text{calc}$ at the end of the run and the 
corresponding $F_\text{calc}$ generated from the full molecule. 

\item Final $R$ factor: 
We calculated the crystallographic R factor (Giacovazzo 
et al., 2002) at the end of the run.

\item  Count error: We compared the integrated real-space electron density 
generated from a run result with the true number of electron as identified 
in the pdb file, on an atom-by-atom basis. The integration was performed 
around each atom over a sphere with a radius that was 1.5 times the grid 
spacing. The figure-of-merit is the RMS error.
\end{enumerate}

Of all these measures, the final $R$ factor was the least useful to assess the 
quality of the reconstructed image. The $R$ factor tends to be lower for a 
small number of entries in the observation file since there are fewer 
equations to satisfy during the reconstruction. In such a case, a visual 
inspection shows that the reconstructed electron density may have little 
resemblance to the 15 carbon atoms. However, there was a good correlation 
among the other three measures. The solutions looked correct when the phase 
difference between solution and full $F_\text{calc}$ was less than 20\r{}, the 
count error between solution and pdb model was 0.2 or lower (out of 6), and 
the RMS distance measure was less than 0.2. 

\subsection{Overcoming Inaccuracies due to Tri-Linear Interpolation Through 
Oversampling}

For computational purposes, we placed the (unknown) molecule or particle 
into a fictitious unit cell that is larger than the molecule, and calculated 
the structure factors on an integer grid in the large unit cell, oversampled 
by the same ratio: the size of the large cell to the size of the molecule. 
We then calculated the structure factor at fractional \textbf{H} from the 
structure factor at integer \textbf{h} using tri-linear interpolation. The 
interpolation error becomes smaller when larger unit cells are used (at the 
expense of computation time). We studied the effect of the unit cell size on 
convergence of \textsc{Speden}.

In the initial tests, the unit cell was the same size as the original 
molecule, and the starting (known) part of the molecule consisted of the 
full molecule. When the molecule consisted of atoms on grid points and the 
$F_\text{obs}$ files had integer $\bm H (i)$, unsurprisingly \textsc{Speden} 
converged, as did \textsc{Eden} on the same data set. However, we found that 
\textsc{Speden} did not converge to a unique solution if either the atoms 
were not on grid points or when the $F_\text{obs}$ file had non-integer 
$\bm{H}(i)$, or both, since the solution meandered in real space.

In subsequent tests, we generated larger unit cells and applied a target 
over the empty part of the unit cell, in an attempt to restrain the 
meandering problem. We embedded the molecule in a cell that was 2 or 3 times 
greater in each dimension. An empty target was used that essentially covered 
the empty 7/8-th or 26/27-th of the unit cell, respectively. We applied a 
high relative weight for this empty target, and we still used the full 
molecule as a starting position. We found that both the larger cell and the 
empty target are of critical value in enabling \textsc{Speden} to converge 
to the correct solution. Comparing the 2-fold larger unit cell versus the 
3-fold unit cell, there was a great improvement in the latter case. These 
results show that for tri-linear interpolation, it is adequate to use a unit 
cell that is 3 times greater in each dimension. We expect that more 
sophisticated interpolation algorithms should allow using smaller unit 
cells.

\subsection{Dependence of Reconstruction on the Quantity of Input Data}

A three-fold enlarged unit cell increases the number of unknown amplitudes 
of the Gaussians, n(p), by a factor of 27. In principle, the emptiness of 
the volume around the molecule restrains the effective number of independent 
unknowns. Nevertheless, if the number of equations, which is given by the 
number of entries in the Fobs file, is not increased, it is easy for the 
solver to ``hide'' electrons among the large number of unknowns in the 
system, even when the empty target constraint is used. In fact we found that 
when we compared the final Fcalc from Solve against the starting Fcalc, on 
the one hand, and the correct Fcalc on the other, Solve's final Fcalc was 
closer to the starting one than to the correct one. In other words, the cost 
function in reciprocal space was not a sufficiently strong constraint in 
SPEDEN's algorithm, for this synthetic problem. In a similar real case, more 
experimental diffraction patterns need to be collected in order that SPEDEN 
would be able to find the corresponding image without other information.

\subsection{Recovery of Missing Atoms}

In the next set of synthetic tests, we attempted to recover missing 
information by starting from a partial model that contained less than the 
full complement of 15 atoms. In these simulations, we used 
$F_\text{obs}$ files 
with non-integer $\bm{H}(i)$, a three-fold enlarged unit cell, an empty 
target or a phase extension target, and randomly positioned atoms. 

We found that a low-resolution spatial target significantly helps 
\textsc{Speden} to converge. Figure 1 (a) shows the results of the 
comparison of the reconstructed image with the electron density from the 
full pdb file when a phase extension target is used. The phase extension 
target was calculated at a resolution of 10 {\AA}. We found that it was 
generally possible to recover 5, 10, or even all 15 of the atoms. Please 
note that the amount of information in such a phase extension target is 
(4{\AA}/10{\AA})$^{3} \quad  \approx $ 6{\%} of the information in the perfect 
solution.

It was more difficult to reconstruct the original electron density when we 
used an empty target, as shown in Figure 1 (b). S\textsc{peden} was able to 
recover 5 out of the 15 atoms, but did not converge when 10 atoms were 
unknown. Perhaps surprisingly, the case where there was no starting model at 
all (0 atoms known) did consistently better than those cases where a partial 
model was given as a starting condition. 

\subsection{Effect of Non-Uniform Sampling on Recovery}

In this set of synthetic tests, we addressed the question of how difficult 
it is to recover the molecule from a non-uniform set of samples in 
reciprocal space, similar to real data sets, and how the results compare 
with the reconstruction from a uniformly-sampled data set.

We generated two-dimensional diffraction patterns of the synthetic carbon 
molecule for different particle orientations, corresponding to recorded 
diffraction patterns in a ``real'' experiment. The two-dimensional 
diffraction patterns were linearly interpolated from a three-dimensional 
diffraction pattern; the latter was calculated on an additionally 
double-fine grid over the already triple-sized unit cells, i.e. using a unit 
cell that was a total of six times larger than the molecule in each 
direction. Further refining the grid of the three-dimensional diffraction 
pattern did not alter the results significantly. The three-dimensional 
diffraction pattern, in turn, was the Fourier transform of the gridded 
electron density of the synthetic molecule. 

Two-dimensional patterns do not sample the diffraction space uniformly. The 
sampling density near the center of the diffraction space is much larger 
than the sampling density further away. We used a completeness measure to 
characterize the sampling uniformity. The reciprocal space is divided into 
cells that are 4$\pi $/$a$ by 4$\pi $/$b$ by 4$\pi $/$c$ in size, where $a$, $b$, and $c$ are 
the molecule sizes in each dimension. The completeness then is the ratio of 
cells in reciprocal space that contain at least one measurement over the 
total number of cells. Figure 2 shows the completeness of the input 
observation files as a function of the number of diffraction patterns. Also 
shown in Figure 2 is the number of calculated diffraction intensities.

We then used \textsc{Speden} to recover 7 out of the 15 atoms. The molecule 
was embedded in a unit cell that was three times larger in each dimension, 
and we used an empty solvent target. Figure 3 shows the errors of the 
reconstructed electron density as a function of the number of 
two-dimensional diffraction patterns. The orientations of the diffraction 
patterns were chosen at random, and the calculations were repeated for four 
different molecules. For comparison, also shown in Figure 3 are the errors 
of the electron density of the 8 known atoms (``partial model''), and the 
errors of the reconstructed electron densities when a three-dimensional 
diffraction pattern is used which was oversampled three times (``integer 
hkl'') or six times (``fractional hkl'').

As discussed above, the $R$ factor is not a useful measure to assess the 
quality of the reconstructed image, but the RMS and count errors are better 
measures for the reconstruction quality. Surprisingly, we found that \textit{four} 
two-dimensional patterns are sufficient to reconstruct the electron density 
as well as in the case the full three-dimensional diffraction pattern is 
given. Four two-dimensional patterns have a remarkably low sampling 
completeness of only 14{\%}. Further increasing the completeness or the 
number of observations does not improve the quality of the reconstruction. 
We would like to note, however, that these results could be dependent on the 
choice of the test model, and that for different test models the number of 
required two-dimensional patterns may be larger.

\subsection{Recovery of Two-Dimensional Data}

In the final set of tests, we demonstrate S\textsc{peden}`s ability to 
recover missing information for a two-dimensional configuration of 37 Au 
balls in a plane. We reconstructed the Au balls using (i) a synthetic 
diffraction pattern and (ii) an experimentally obtained diffraction pattern 
as discussed by He et al., 2003. In the following we will discuss both 
cases, starting with the synthetic diffraction data.

The 37 Au balls are arranged in a plane as shown in Figure 4. The 
arrangement of the balls is similar to the experimental case discussed by He 
et al., 2003. The Au balls were 50 nm in diameter. We generated an 
artificial set of ``measurements'' ($\vert F_\text{obs}\vert $) by 
calculating the structure factors ($F_\text{calc})$ to 30 nm resolution and 
deleting the phases. The uncertainty of the measured structure factors, 
$\sigma(\bm{h})$, were chosen according to Equation (\ref{eq10}). We generated 
an initial model by smearing the full $F_\text{calc }$ file to 90 nm, 
and running \textsc{Back} on it. 
The initial model is shown in Figure 5. In Figures 5 --7, we 
only show one plane. We also used this smeared $F_\text{calc}$ to generate a 
low-resolution spatial target as well as an empty target outside the 
molecule. The corresponding weight function is shown in Figure 6. 

We then used \textsc{Speden} to reconstruct the Au balls. As shown in Figure 
7, \textsc{Speden} reconstructed the electron density successfully. However, 
we further found that if we use an empty starting model, S\textsc{peden }has 
difficulties converging to the correct electron density. There are two 
reasons for this behavior. First, without an initial model, the symmetry of 
the system is not broken, and \textsc{Speden} stagnates since the support 
does not distinguish between the object and its centrosymmetric copy. 
Second, the mask and the reconstructed electron density are possibly shifted 
with respect to each other. If the initial model is empty, the position of 
the reconstructed electron density is mostly determined in the early 
iteration of the \textsc{Solve} algorithm and can be partially cut off by 
the solvent. The algorithm has difficulty shifting the result. It is 
necessary to provide information about the location of the electron density 
to a certain degree, for example in the form of a smeared model. Note that 
the GSF algorithms are designed to overcome these problems when there is 
abundant and accurate data available.

We will now discuss the reconstruction of the Au balls using experimental 
data. To generate a starting model, we took the experimental $\vert 
F_\text{obs}\vert $ data along with the phases obtained by He at al., 2003 
using a version of the GSF algorithm, and smeared this data to 80 nm. The 
starting model is shown in Figure 8. We used the same data to generate a 
real space target with a target fraction of 99.7{\%}, shown in Figure 9. 
Similar to the case of the synthetic test data, we chose $\sigma 
$(\textbf{h}) according to Equation (\ref{eq10}). 
We then used \textsc{Speden} to 
reconstruct the Au balls. Figure 10 (a) shows the reconstructed electron 
density, and Figure 10 (b) shows a scanning electron microscope (SEM) 
picture of the sample. We found that S\textsc{peden} reconstructed the 
electron density from the experimental data successfully.

\section{Summary and Conclusions}

In this paper we presented \textsc{Speden}, a method to reconstruct the 
electron density of single particles from their x-ray diffraction patterns, 
using an adaptation of the Holographic Method in crystallography. Unlike 
existing GSF algorithms, \textsc{Speden} minimizes a cost function that 
includes all the constraints of both real space and reciprocal space, by 
varying quantities in real space only, so that the cost evaluation requires 
only a forward transform from real space to reciprocal space. 
S\textsc{peden} finds a local minimum of the cost function using the 
conjugate gradient algorithm. We implemented \textsc{Speden} as a computer 
program, and tested it on synthetic and experimental data. Our initial 
results indicate that \textsc{Speden} works well. 

\ack{Acknowledgments. }
This work was performed under the auspices of the U.S. Department of Energy 
by University of California, Lawrence Livermore National Laboratory under 
Contract W-7405-Eng-48 and DOE Contract DE-AC03-76SF00098 (LBL). SM 
acknowledges funding from the National Science Foundation. The Center for 
Biophotonics, an NSF Science and Technology Center, is managed by the 
University of California, Davis, under Cooperative Agreement No. PHY0120999.

\begin{figure}
\includegraphics[width=8cm]{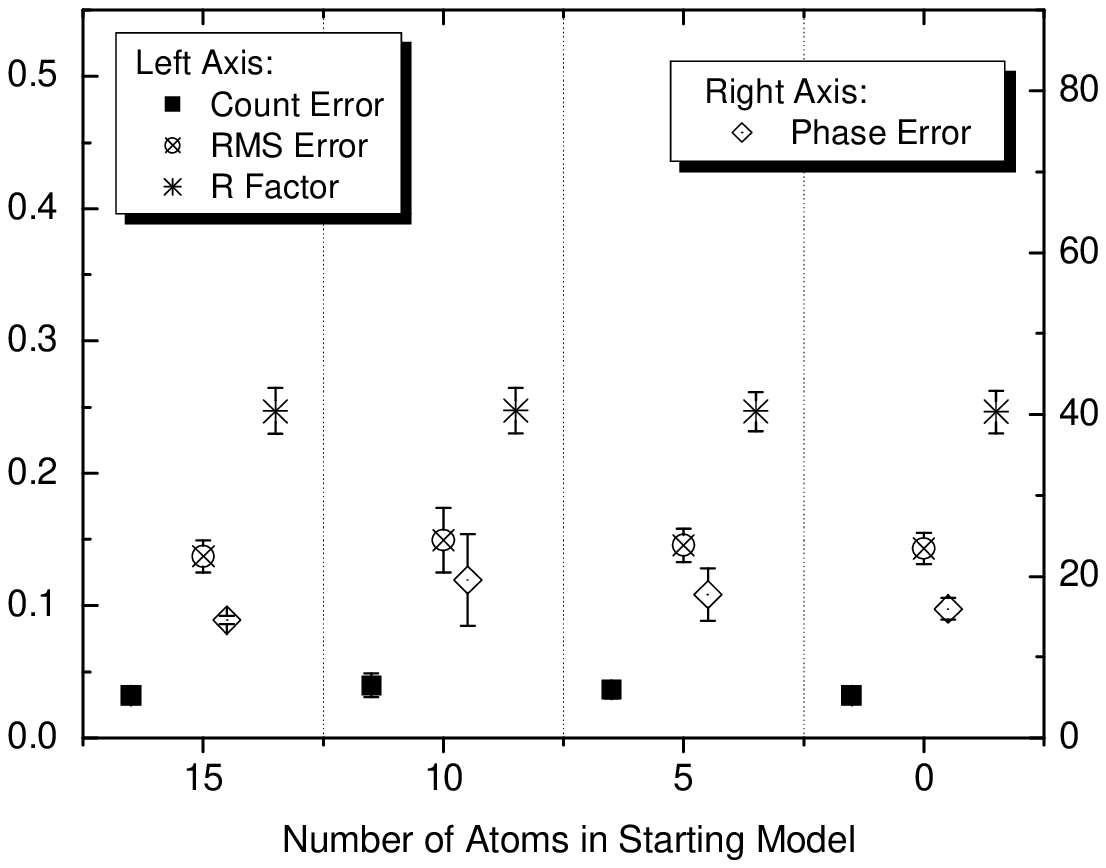}
\includegraphics[width=8cm]{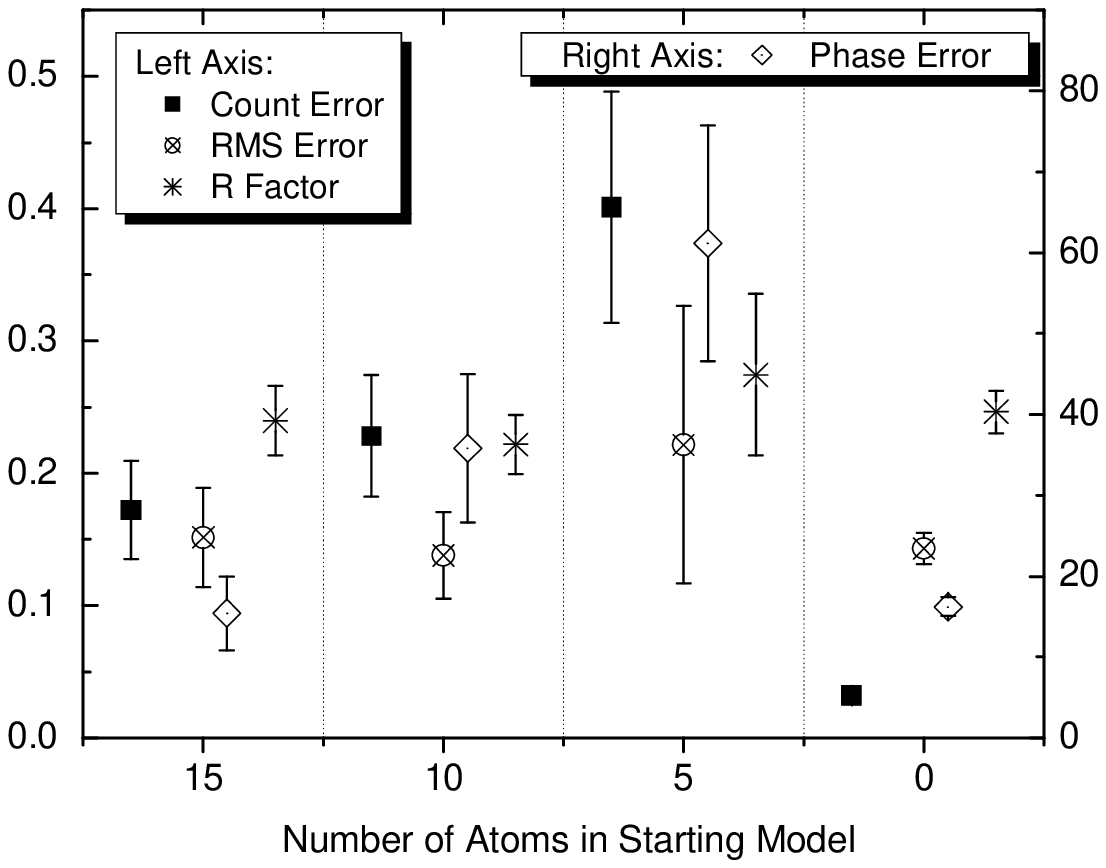}
\caption{(a)
 Results of the comparison of the reconstructed image with the electron 
density from the full pdb file in the cases (a) a phase extension target and 
(b) low-resolution target are used. The phase error is given in units of 
degrees.}
\end{figure}

\begin{figure}
\includegraphics[width=9cm]{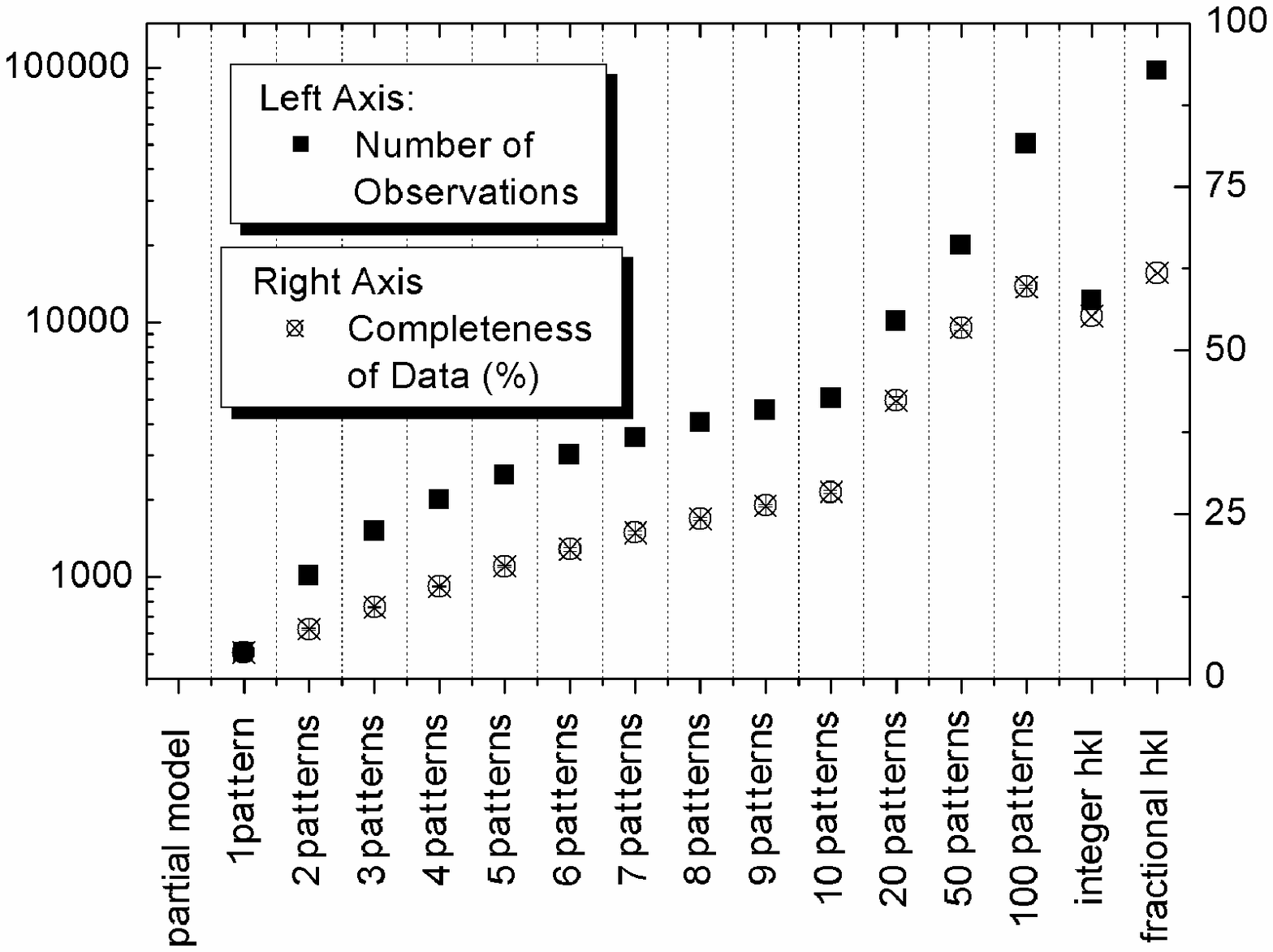}
\caption{
The completeness and length of the input observation files used for the 
calculations shown in Figure 3.}
\end{figure}

\begin{figure}
\includegraphics[width=9cm]{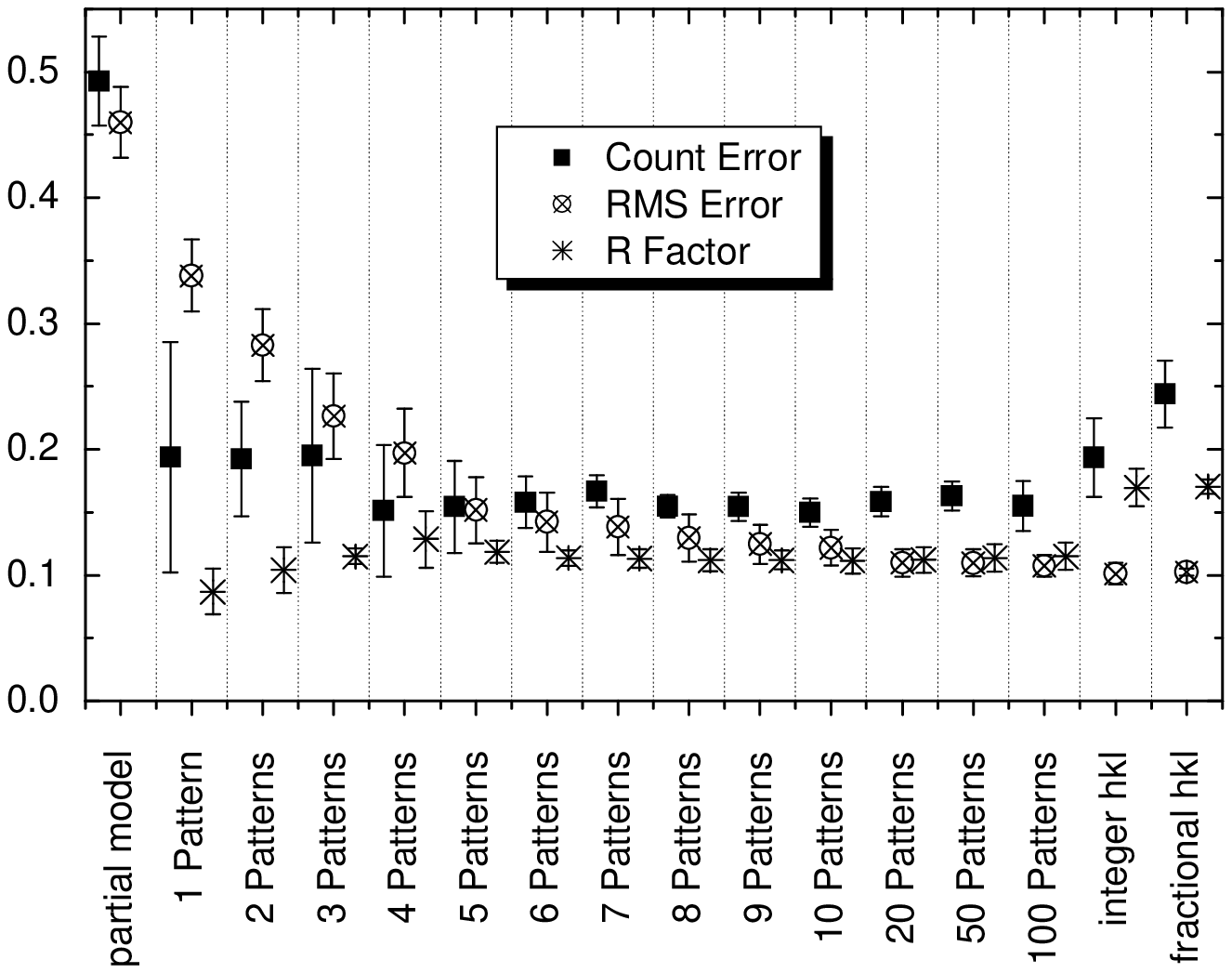}
\caption{
The error in the reconstructed electron density, as a function of the number 
of two-dimensional diffraction patterns. The orientation of the diffraction 
patterns was chosen at random, and the results were repeated for four 
different molecules. Also shown are the error of the electron density of the 
8 known atoms (``partial model''), and the error of the reconstructed 
electron densities for a three-times (``integer hkl'') and six-times 
(``fractional hkl'') in-each-dimension-oversampled three-dimensional 
diffraction pattern.}
\end{figure}

\begin{figure}
\includegraphics[height=8cm]{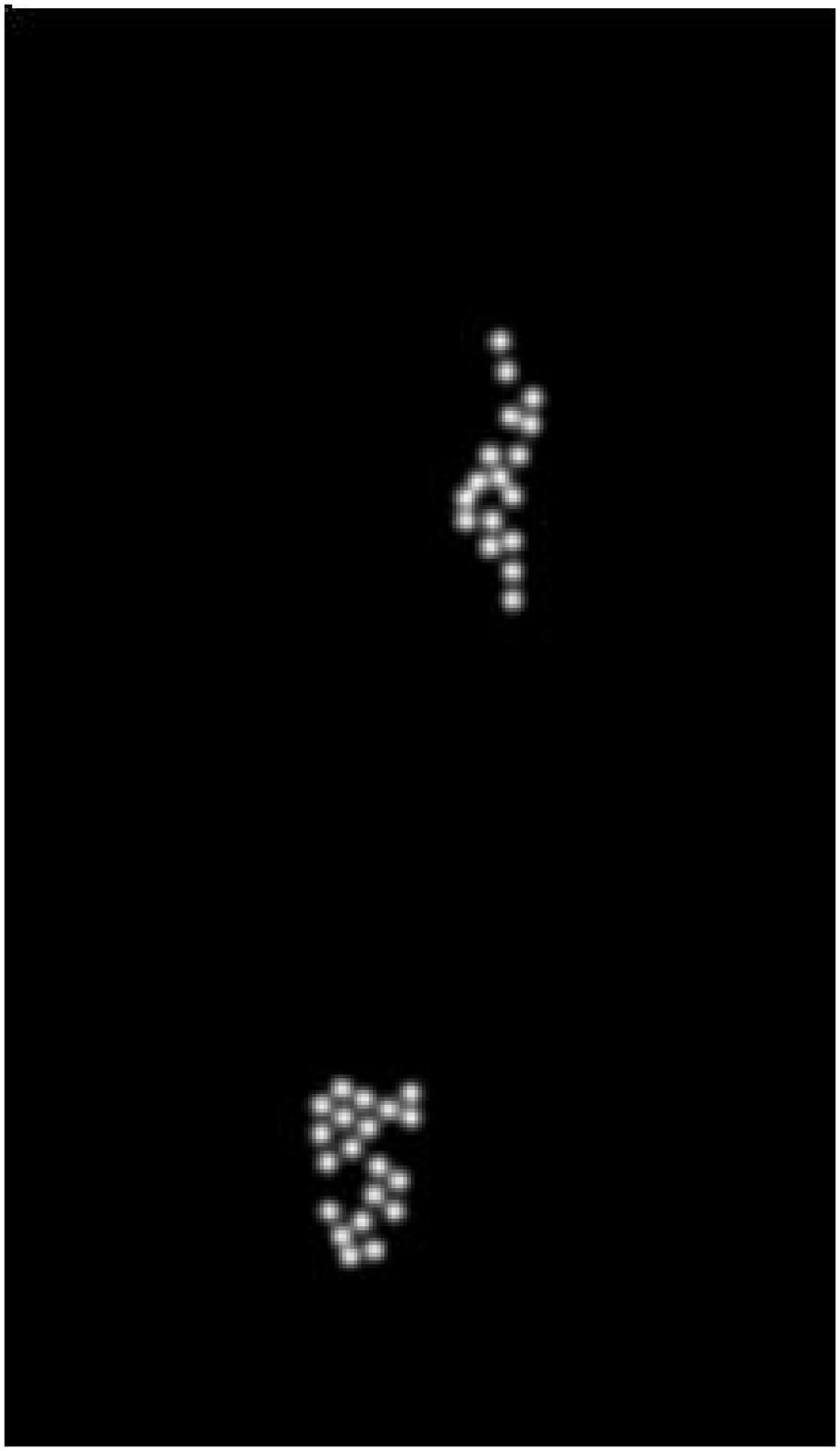}
\caption{
Planar arrangement of 37 Au balls for two-dimensional reconstruction.}
\end{figure}

\begin{figure}
\includegraphics[height=8cm]{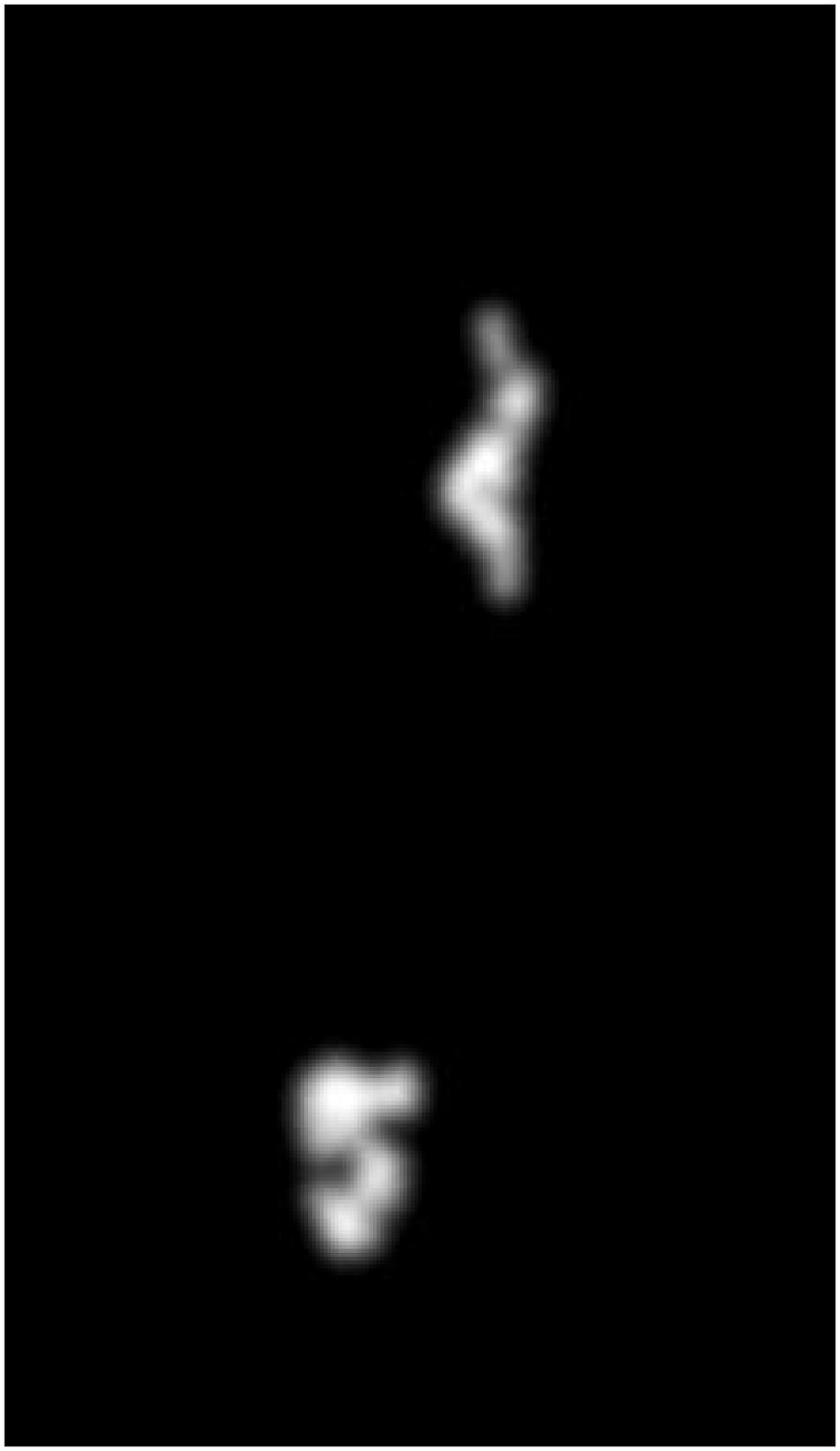}
\caption{
Initial starting model used for two-dimensional reconstruction from 
synthetic data.}
\end{figure}

\begin{figure}
\fbox{\includegraphics[height=8cm]{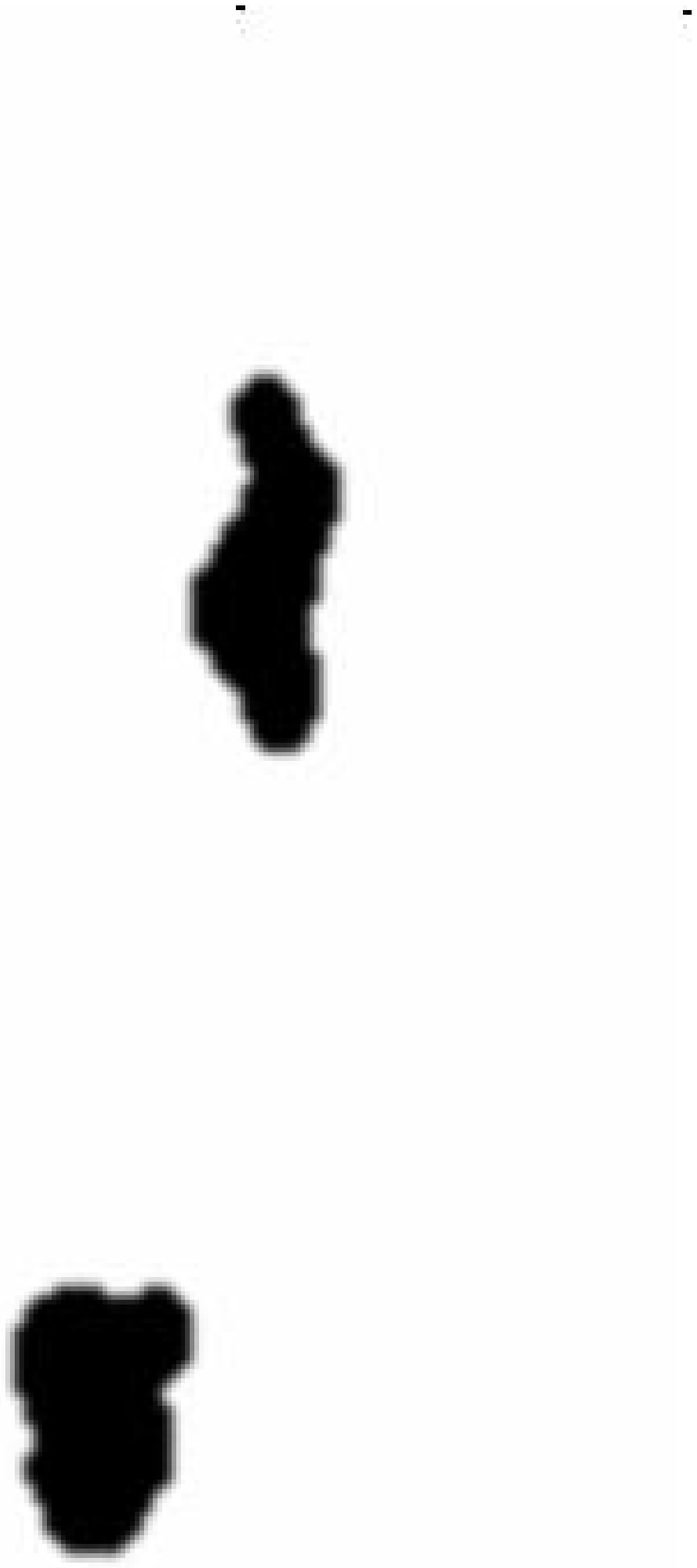}}
\caption{
Weight function used for two-dimensional reconstruction from synthetic data.}
\end{figure}

\begin{figure}
\includegraphics[height=8cm]{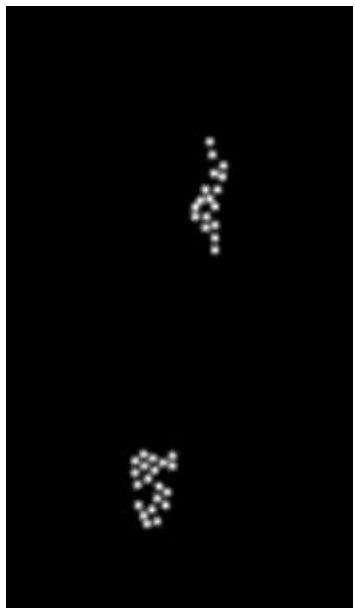}
\caption{
Reconstructed electron density using synthetic data.}
\end{figure}

\begin{figure}
\includegraphics[height=8cm]{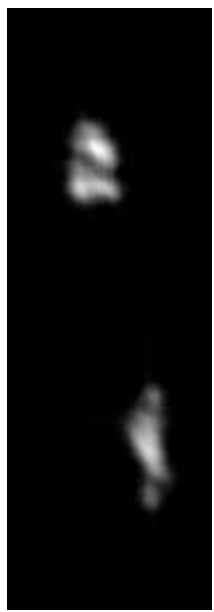}
\caption{
Initial starting model for two-dimensional reconstruction from experimental 
data.}
\end{figure}

\begin{figure}
\fbox{\includegraphics[height=8cm]{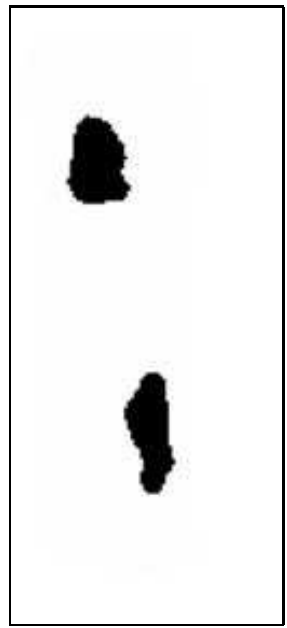}}
\caption{
Weight function used for two-dimensional reconstruction from experimental 
data.}
\end{figure}

\begin{figure}
\includegraphics[height=8cm]{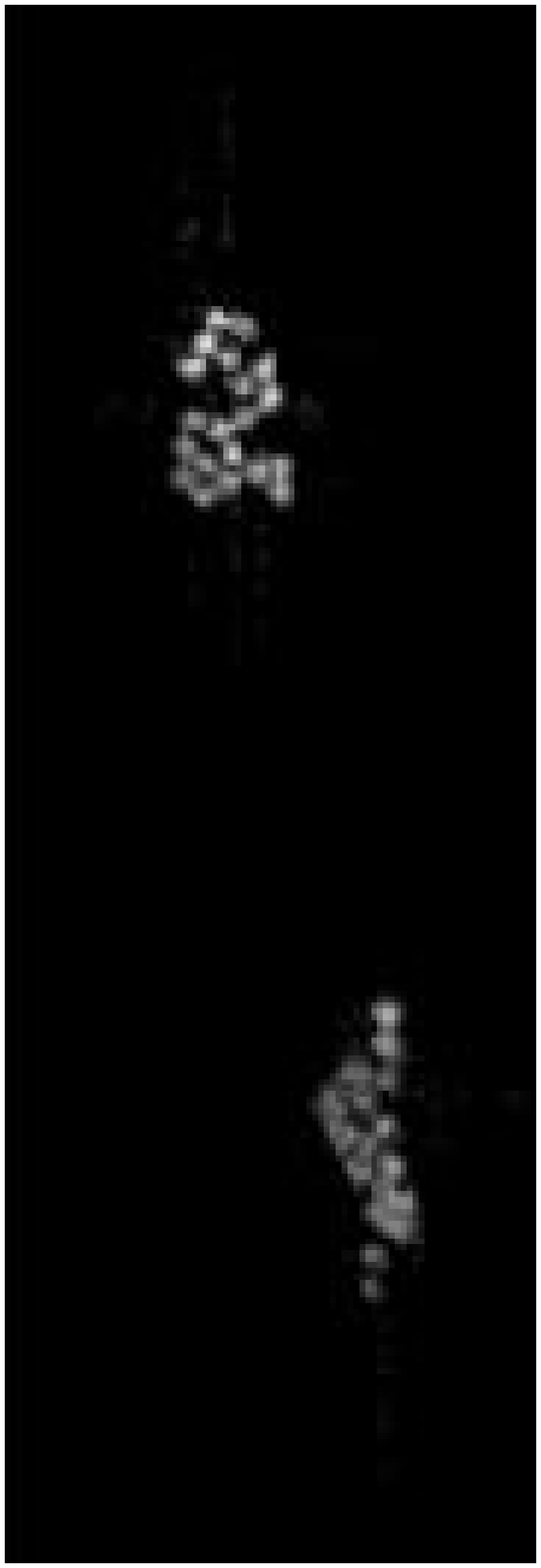}
\includegraphics[height=8cm]{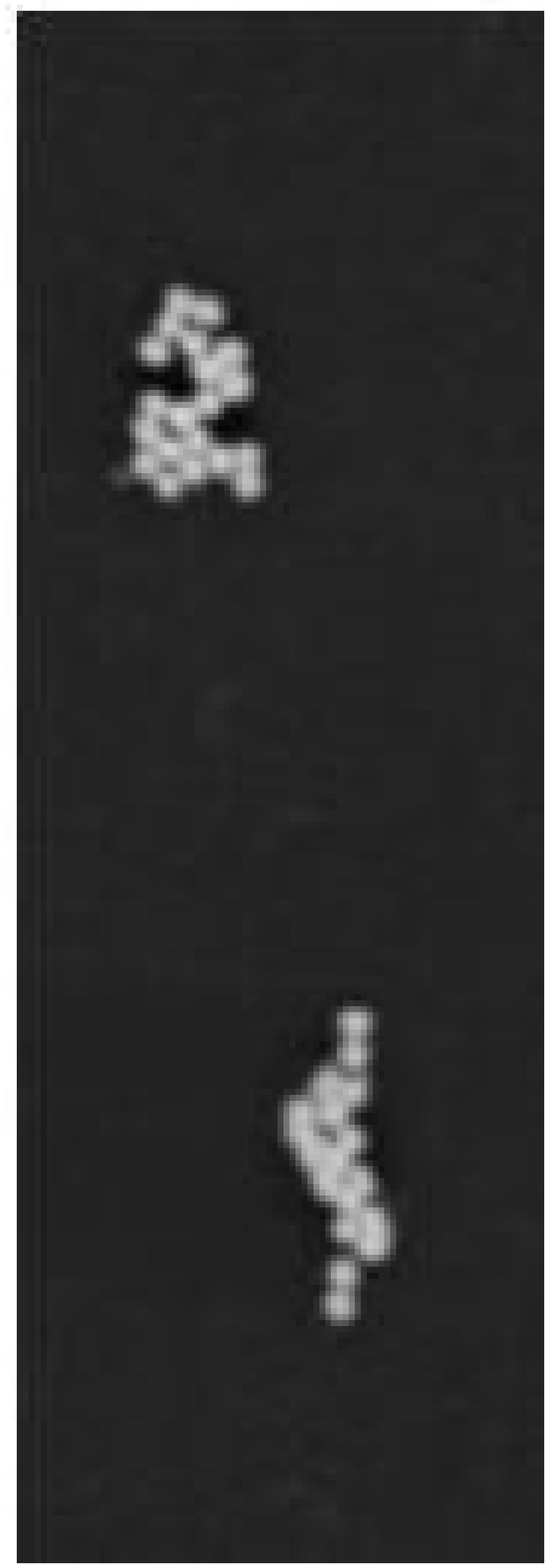}
\caption{
(a) Reconstructed electron density using experimental data. (b) SEM Image of 
Au balls.}
\end{figure}

\end{document}